\documentclass[preprint,12pt,a4paper]{elsarticle}
\usepackage{graphics}

\journal{Nuclear Physics A}

\begin{document}

\begin{frontmatter}

\title{The strength of beta-decays to the continuum}

\author[aar]{K. Riisager\corref{cor1}}
\address[aar]{Department of Physics and Astronomy, Aarhus University,
 DK-8000 Aarhus C, Denmark}
\cortext[cor1]{E-mail: kvr@phys.au.dk}

%\begin{document}
%
%\maketitle

\begin{abstract}
  The beta-strength in beta-delayed particle decays has up to now been
  defined in a somewhat ad hoc manner that depends on the decay
  mechanism. A simple, consistent definition is presented that fulfils
  the beta strength sum rules.  Special consideration is given to the
  modifications needed when employing R-matrix fits to data. As an
  example the $^{11}$Be($\beta$p) decay is investigated through simple
  models.
\end{abstract}

\end{frontmatter}

\section{Motivation}

Close to the beta-stability line all beta-decays will populate
particle-bound states, i.e.\ states that are longlived (stable, beta
or gamma decaying) and therefore have narrow widths, less than 1 keV.
Moving towards the driplines a larger and larger fraction of
beta-decays will feed states that are embedded in the continuum.  A
general overview of the physics changes this brings about can be found
in recent reviews \cite{Bla08,Pfu12}. Close to the dripline 
beta-delayed particle emission can become the dominating decay mode
and the question of how beta strength is assigned to transitions
to unbound levels becomes important.
This has been discussed at several instances, e.g.\
\cite{Bar69,Bar88,Nym90,Bar96}, the aim of this paper is to provide a
consistent answer that is independent of the mechanism for the
particle emission.
Quite apart from the conceptual interest this also
has a very practical implication for the way the total beta strength
is calculated: as remarked earlier \cite{Nym90} current approaches
give a
strength corresponding to decays in an energy region that is proportional
to $\int (ft)^{-1} \mathrm{d}E$ for decays going directly to the continuum versus 
$(\int ft \,\mathrm{d}E)^{-1}$ for decays through a resonance.
 
To simplify the notation I shall mainly consider Gamow-Teller
transitions.  For transitions to bound states the decay rate is $w =
\ln 2 \frac{g_A^2}{K} f(Q-E) B_{GT}$, where $f$ is the phase space
factor, $K = 2\ln 2 \pi^3\hbar^7 /(m_e^5c^4)$ ($m_e$ being the electron
mass), the beta strength $B_{GT}$  is given by the reduced matrix 
element squared $|\langle f| \beta_{\pm} |i\rangle|^2$
and the weak interaction constant $g_A$ is factorized out explicitly
from the operator $\beta_{\pm}$ that flips spin and isospin.

The basic suggestion of this paper is to define the beta strength for
final unbound states so that the following expression holds for the
decay rate:
\begin{equation}   \label{eq:def}
    w(E) \mathrm{d}E = \ln 2 \frac{g_A^2}{K} f(Q-E) B_{GT}(E) \mathrm{d}E  \;, 
\end{equation}
where there is an implicit sum over all final states with the same
$E$. This definition is in principle experimentally simple to
implement, but can be more complex to use theoretically since it does
not distinguish between different cases such as isolated resonances in
the continuum, interfering resonances, one or several decay channels
etc. Essentially one takes out the lepton part ($Q-E$ is the energy
going to the beta particle and the neutrino), so it is an attempt to
separate the weak interaction part (the ``incoming channel'') and the
strong interaction part (the ``outgoing channel''). A complete
separation is not possible unless each decay goes through one and only
one intermediate state. As explained in detail later
different definitions of $B_{GT}$ have been employed in earlier
papers.

Section 2 presents an argument based on the Gamow-Teller sum rule for
why the above suggestion is appropriate, and the two following
sections compares the definition to existing frameworks. Section 3
looks in detail on beta decays going directly to continuum states and
how they have been treated theoretically so far. Section 4 deals with
the treatment of decays through intermediate resonances as done in the
R-matrix formalism and how this can be modified to be consistent with
the proposed definition. Section 5 presents the conclusions and the
appendix gives more mathematical details relevant for the R-matrix
treatment.

\section{Beta strength sum rule}

The Gamow-Teller sum-rule is very useful for beta decay studies.  It
gives a natural scale for $B_{GT}$ for a given decay and is derived by
using the completeness relation for rewriting the summed strength for
an initial state $|i\rangle$ as
\[  S_{\pm} (GT) = \sum_f |\langle f| \beta_{\pm} |i\rangle|^2
     =  \sum_f \langle i | \beta^{\dagger}_{\pm}|f\rangle\langle
     f|\beta_{\pm}|i\rangle  =
     \langle i|\beta^{\dagger}_{\pm}\beta_{\pm} |i\rangle 
\]
and by evaluating the commutation relations of the beta operators
one gets
\[  S_-(GT) - S_+(GT) = 3 (N-Z)  \;.
\]
In this standard derivation of the sum-rule one implicitly assumes a
discrete set of final states each with beta strength $B(GT)_f =
|\langle f| \beta_{\pm} |i\rangle|^2$, this must of course be changed
when significant contributions come also from continuum states.

A pragmatic way of proceeding that shall be explored later in section
\ref{sec:dcdd} is to use as a first step a discretized continuum by
imposing a finite (but large) quantization volume. By construction
the rules for a discrete spectum applies and the sum rule is
unchanged. In the continuum limit of increasing quantization volume
one would naturally obtain equation (\ref{eq:def}) and the
Gamow-Teller strength will obey the sum rule. Calculations of
continuum spectra that proceed by this route will be safe, but other
approaches are possible that throw more light on the intricacies of
the continuum.

%[ Then do the formal treatment (in this way the normalization is no problem !):]

A more formal treatment of the question of how to formulate the
completeness relation including continuum states was given by Berggren
and collaborators \cite{Ber68,Ber93}. 
% By far most beta-delayed processes are described as proceeding through
%resonances in the daughter nucleus. Assuming this decay mechanism will
%often be the obvious choice from the
%resulting spectra (when the decay produces narrow peaks) or it may be
%needed in order to perform meaningful comparisons to theoretical
%calculations. Formally, it was shown by Berggren and collaborators
% how to change between a description in terms of
%continuum and one in terms of resonances. 
With a careful definition of the continuum wavefunctions one can derive
general sum rules \cite{Ber73,Rom75} where for our specific case the
sum over discrete states is replaced by a sum over bound states and an
integral over all (real values of the) momenta in the continuum
\begin{equation}   \label{eq:Berg}
   \sum B_{GT} + \int B_{GT}(k) \mathrm{d}\vec{k}\;.
\end{equation}
%in analogy to our eq.\ (2), this incidentally shows that our $B_{GT}$
%will fulfill the Gamow-Teller sum rule. 
Berggren further showed how one by allowing complex momenta $k$ and
modifying the contour of integration in eq.\ (\ref{eq:Berg}) could
extend the sum over bound states to include also contributions from
resonance states. Conceptually this gives the crucial insight that
even though the physical decay mechanism may favour the description in
terms of resonances or the one in terms of continuum transitions we
are in principle at liberty to use both (or, in the general case, a
mixture).  There are two important points to note: first that even
when all physical resonances are included in the sum there may remain
a small continuum contribution, secondly that in practical
implementations one may encounter non-positive contributions from
individual terms in the sum as shown explicitly in \cite{Myo98} for
the corresponding case of an electric dipole. The resonances that
emerge in this framework can therefore not be replaced by or simply
identified with the resonances occuring e.g.\ in the R-matrix
framework.

If one in equation (\ref{eq:Berg}) integrates $B_{GT}(k)$ over all
momenta corresponding to the same energy $E$ one obtains the
$B_{GT}(E)$ from above. It is therefore possible to consistently
define Gamow-Teller strength in the ``pure continuum'' so that the sum
rule is maintained. If one wishes to assign strength to a specific
resonance this can be done, but there is in principle a risk of
obtaining non-positive values. The question of when continuum
contributions will remain important is treated in \cite{Ber73,Rom75}.
%[References for investigations of sum rules based on Berggren
%expansion: Berggren \cite{Ber73}, Romo \cite{Rom75} that also gives
%qualitative answers to where the continuum contributions are of
%importance, a use of the sum rules in checks of neutrino scattering
%\cite{Civ08}, the use of sum rules to include continuum contributions
%in electromagnetic transitions \cite{Myo97}. Also other papers ?]

\section{Decay directly to continuum}
The Berggren approach is being implemented in nuclear structure
calculations via the Gamow Shell Model \cite{Mic09}, but has so far
not been applied to beta-delayed particle emission. Calculations of
beta decays directly to a continuum state $|k\rangle$ have been made
within several approaches with
different conventions for the normalization of
continuum states and correspondingly different choices 
for the normalization of the reduced matrix element
$B_{GT}(k)$: in \cite{Rii90} the continuum was discretized in a large
volume and the wavefunction normalized to one particle per volume, in
other calculations, see e.g.\ \cite{Zhu93,Bay94,Tur06}, the wavefunctions
at large radii become scattering wavefunctions.
When calculating the decay rate as a function of the
continuum energy $\mathrm{d}w(E)$ one must sum over all states with
the same energy $E$. An explicit ``phase space factor'' for the
outgoing particle should therefore be included, a factor
that of course also depends on the chosen normalization
thereby
bringing some confusion to the notation\footnote{In fact, earlier
  papers use the notation $B_{GT}(E)$ for the present $B_{GT}(k)$.}.  
The conversion to the present definition is for \cite{Rii90}
\[   B_{GT}(E) = B_{GT}^H(E) \frac{k}{2\pi^2} \frac{mc^2}{(\hbar c)^2}  \;,
\]
where $k$ and $m$ are the momentum and mass of the outgoing
particle. For \cite{Zhu93} one has
\[   B_{GT}(E) = B_{GT}^Z(E) \frac{1}{2\pi^2 v}  \;,
\]
where $v$ is the velocity of the outgoing particle. For \cite{Tur06}
one has
\[   B_{GT}(E) = B_{GT}^B(E) \left( \frac{g_V}{g_A} \right)^2 \frac{2}{\pi\hbar v}  \;,
\]
where the extra ratio of coupling constants is due to a different
convention that includes them in the definition of $B_{GT}$.

From the previous section it follows that calculations of beta decays
going directly to continuum states should essentially automatically
fulfil the Gamow-Teller sum rule. The main point in the present
definition, equation (\ref{eq:def}), is for this case only a redefinition of
$B_{GT}(E)$ as a sum of all $B_{GT}(k)$ so that the calculation
dependent ``phase space factor'' is not included in the strength
definition. This makes comparisons between calculations and between
experiment and theory easier.

Up to now the decays that have been described as going directly to 
continuum states are some of the decay channels of halo nuclei
\cite{Pfu12}.  In the
specific case of beta-delayed deuteron decays of two-neutron halo
nuclei this is the standard assumption in most theoretical
descriptions of the process (based on the picture of the two halo
neutrons decaying ``remotely'' into a deuteron), see e.g.
\cite{Tur06} for the most recent calculation of this decay mode in
$^6$He. However, a description within the R-matrix approach has also
been done \cite{Bar94} and more experimental data may be needed in
order to settle whether direct decays is the only reasonable
description.

\section{Sequential decay}
The case of decays through resonances is considerably more complex, in
particular for broad resonances where the $f$-factor changes
significantly across the level and where interference may play a
role. This case is typically analysed with the R-matrix formalism
\cite{Lan58,Des10} that allows adjusting level parameters to better
fit experimental data.
Before going into technical details it may be useful to remind why a
resonance description is used at all. It is the natural description
when there are narrow lines in the experimental spectrum, but it is
also of interest more generally since
a resonance description summarizes much information
into a few numbers. If a few resonances can describe all the
structure in a spectrum it gives an economical description that
furthermore can be extrapolated (with caution) to neighbouring regions
that may be harder to access experimentally. 

The R-matrix approach (or an equivalent framework, see \cite{Rob75}
for an overview of theoretical approaches that have been used to
describe resonance reactions) is
essential if there is strong coupling to the continuum or if
resonances overlap so that interference occurs, some examples from the
light nuclei are the decays of $^8$B, $^{12}$N, $^{17}$Ne and $^{18}$N.
%$^8$He, $^{16}$N, $^{20}$Na ??]
Appendix A contains a more detailed exposition of the R-matrix
formalism for beta-delayed decays.
I shall mainly consider the single level, singel channel case where,
as shown in \cite{Bar96}, the decay rate is 
\[   \mathrm{d}w = \ln 2
     \frac{g_A^2}{K} f(Q-E) B_{GT}^R \frac{\rho(E)}{\pi} \mathrm{d}E 
\]
where the size of the
beta strength parameter $B_{GT}^R$ (essentially the square of the
parameter $g$ in \cite{Bar88}) depends explicitly on the
normalization of the line shape that is given by
\begin{equation}  \label{eq:rho}
  \rho_{\lambda}(E) = \frac{P(E)\gamma^2_{\lambda}}{
    [E_{\lambda}-E+(S(E)-B) \gamma^2_{\lambda}]^2 +
    [P(E)\gamma^2_{\lambda}]^2} \;.
\end{equation}
Here $P,S$ and $B$ are the penetrability, shift function and boundary
parameter and $E_{\lambda}$ and $\gamma_{\lambda}$ are the level
energy and width parameters. By comparison to the continuum
description one sees that $B_{GT}^R$ will give the summed beta
strength for an isolated level if the integral of $\rho(E)$ is $\pi$.
% and the R-matrix expression gives a good fit to the data
If the integral differs from $\pi$ the basic suggestion of the present paper
is that the strength derived from the continuum description is the
correct one (it leaves the sum rule unchanged) and is related to the R-matrix
parameter $B_{GT}^R$ through 
\begin{equation}  \label{eq:condef}
    B_{GT}(E) = B_{GT}^R \rho(E)/ \pi
\end{equation}
so that the integrated strength of the decay through a specific
isolated level is $B_{GT}^R \int \rho(E) \mathrm{d}E/\pi$.

A very similar correction has been applied by Barker \cite{Bar69,Bar88,Bar96},
who for narrow levels approximates 
$  \int \rho(E)\mathrm{d}E =
  \pi/(1+\gamma^2_{\lambda} \mathrm{d}S/\mathrm{d}E )$
where the derivative is evaluated at $E_{\lambda}$.
This question is analysed in more detail in Appendix A where the
limitations of the approximation are exposed.
There is no general unique prescription that in a simple manner will
give the total beta strength corresponding to a level.  Furthermore,
if one tries to determine the total strength by performing the
integral $\int \rho(E) \mathrm{d}E$, the contribution to the integral
above an energy $E_h$ will be proportional to
$\gamma^2_{\lambda}/\sqrt{E_h}$ and therefore be potentially large
for wide levels. It is not obvious that this contribution at high
energies is physically relevant and therefore not obvious which
upper integration limit should be used. The best one can do is to
employ equation (\ref{eq:condef}) and e.g.\ determine the strength for
decays through a specific level in a given energy range.

The extension from the single-channel, single-level case to the more
general situation will not give qualitative changes in the
picture. Numerically, the interference that enters in the multi-level
case redistributes the beta strength rather than changing its total
value. (It seems also to diminish the dependence on $E_h$ mentioned
above.)  In any case, if interference effects are large the whole
procedure of attributing beta strength to each individual level may be
questioned. The beta strength to a given level cannot be extracted
immediately from a spectrum. If one in such cases choses to quote a
$B_{GT}$ (or, equivalently, an $ft$-value) the price to pay is that the
sum rule is no longer valid and that an evaluation of total strength
directly from the spectrum will give a different result (that fullfils
the sum-rule).

A pragmatic way to extract beta strength when fitting with
the R-matrix formalism is the following: if the resonances are narrow
and isolated one can normally use the same procedure as for bound
states, except when the variation of the $f$-factor across the level is
substantial. In the latter, and other more complex cases, one can
either switch to using eq.\ (\ref{eq:def}) or equivalently use
equation (\ref{eq:condef})  and calculate explicitly
the integral of $\rho(E)$ or the corresponding integrals for the
multi-channel, multi-level cases given in \cite{Bar88}. 
Barker uses the Q-value as the upper limit for the integration range
(this would correspond to including only the observed strength within
the energetically available window), but if the choice has any effect
it must be carefully stated.

\subsection{A model case: $^{11}$Be($\beta$p)} \label{sec:dcdd} 
The general results will now be exemplified via a simple tractable
case, namely the beta-delayed proton emission from $^{11}$Be. This
decay mode should be similar to the beta-delayed deuteron decays from
two-neutron halo nuclei mentioned above, but is conceptually
simpler. A recent paper \cite{Bor13} contains more details on this
decay with references to the literature. The model considered here is
too simple to be applied immediately to the decay and e.g.\ does not
consider isospin nor decays of core nucleons. Nevertheless, it will
serve as a useful illustration.

The basic assumption in this discretized continuum direct decay model
(DCDD) is that the initial state is an s-wave neutron in the potential
given by $^{10}$Be that is assumed to be inert. The final states
are continuum wave functions of a proton in an s-wave in the combined
Coulomb and nuclear potential from $^{10}$Be. The Gamow-Teller
operator simply converts the halo neutron into a proton (the spin operator
will not change the physics) so the matrix element reduces to the
overlap between the two wavefunctions. Fermi transitions are assumed
to go mainly to the isobaric analogue state and are therefore not
included. The final wavefunctions are found as the discrete set of
positive energy solutions in a finite volume. The ``energy
resolution'' given by the differences in level energies decrease as
the radius of the quantization volume is increased; in the
calculations the radius of the volume varied between 400 fm and 4000 fm.

The strong potential between the core and the nucleons is taken as a
square well of radius 4.0 fm. This gives an appropriate halo
wavefunction for the $^{11}$Be ground state when using wavefunctions
with one node inside the potential (if wavefunctions with no node are
used, the potential radius should be reduced to 3.5 fm). For the
initial state the potential strength is adjusted to 33.819 MeV to fit
the known $^{11}$Be neutron separation energy \cite{mas12} of
501.64(25) keV.  For the final state a square well is used up to 4 fm
and a pure Coulomb potential for radii beyond this.  The structure of
the solutions depend on the well depth used in the final state. For
most values one obtains small overlaps with wavefunctions within the
280.7 keV window open for $\beta$p decay and a featureless spectrum. A
very similar result was obtained in the more sophisticated two-body
calculations of the decay in \cite{Bay11}. This is the
``non-resonant'' regime with nothing conspicuous appearing in the
calculated decay spectrum.

\begin{figure}[thb]
\centering
     \includegraphics[width=8.cm,clip]{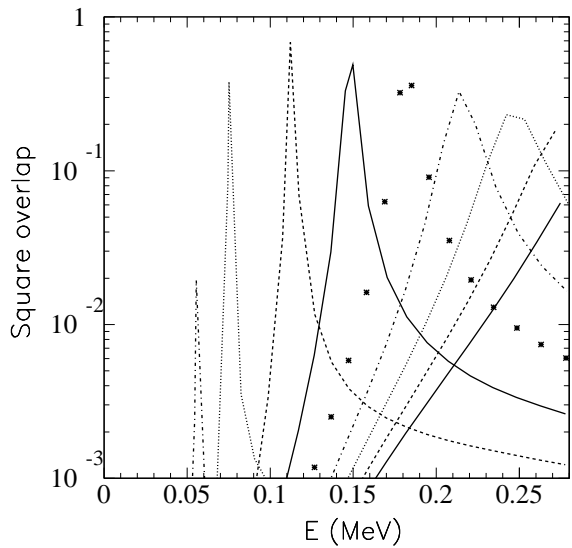} 
     \caption{The squared overlap are plotted for each final state
       wavefunction in the p+$^{10}$Be system calculated with a square
       well of radius 4.0 fm and a confining
       radius at 1000 fm. The curves are for potential depths (in MeV)
       of 33.85, 33.8, 33.7, 33.6, 33.5, 33.4, 33.3, 33.2 and 33.1.
       The initial single-particle wave function is the same for all cases
       and corresponds to a separation energy of 501.64 keV. All
       wavefunctions have one node within the potential well.}
\label{fig:over_V} 
\end{figure}

When the depth of the final square well potential is in the range
33--34 MeV one obtains significantly higher overlaps (in this range
the final state wavefunctions have one node inside the potential,
significant overlaps are also obtained in limited regions where
wavefunctions have no or two nodes inside the potential). The obtained
overlaps are shown in figure \ref{fig:over_V} for a 1000 fm confining
radius. One observes a ``resonance'' inside the small window with a
width that becomes smaller for lower resonance positions. This
behaviour is in contrast to the one observed if one puts the Coulomb
potential to zero in the final state (as if the final nucleons were
neutrons) in which case there is a broad maximum in the overlaps.
%for potential depths around 3.5 MeV.  
A similar behaviour to the one seen
here appears in the calculations of beta-delayed deuteron emission in
\cite{Zhu95} where the final state spectra generally are broad and
almost featureless but display a low energy peak for small ranges of
the potential depth. 

It may be of interest to briefly compare the found resonances with the
famous case of the $^8$Be ground state. Comparing the Schr\"{o}dinger
equations in the tunneling region for the $\alpha+\alpha$ case and for
p+$^{10}$Be one finds that scaling the radius in the latter case up by
the ratio of reduced masses in the two systems and the energies down by
the same factor, one obtains exactly the same equation. In other words,
the $^8$Be ground state corresponds to a proton resonance
at 42 keV. Higher resonance energies correspond to systems
more unstable than $^8$Be.
In a similar way, when the p+$^{10}$Be system is compared to the
d+$^9$Li one, if the proton radial distances are scaled by a factor
27/20 and the proton energy by a factor 81/80 one obtains exactly the
same Schr\"{o}dinger equation for the two systems in the region below
the Coulomb barrier. I.e., the energy scaling factor is here very
close to one.

\begin{figure}[thb]
\centering
     \includegraphics[width=8.cm,clip]{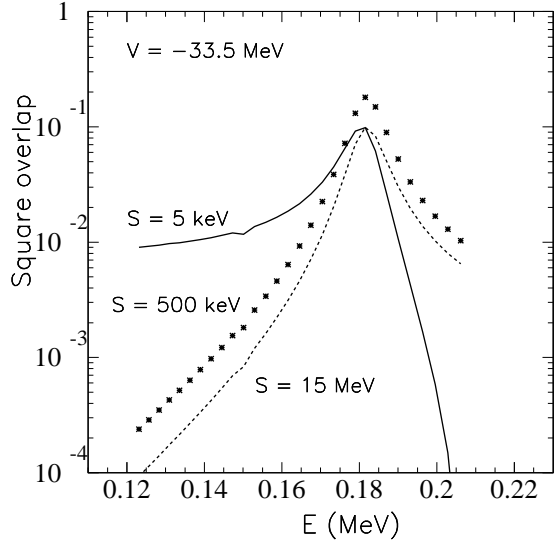} 
     \caption{The squared overlap are plotted for each final state
       wavefunction in the p+$^{10}$Be system calculated with a square
       well of radius 4.0 fm and depth 33.5 MeV and a confining
       radius at 4000 fm. The three curves are results for initial
       single-particle wave functions with different neutron
       separation energies: 5 keV (solid line), 500 keV (a star for
       each wavefunction) and 15 MeV (dashed line).}
\label{fig:over_E} 
\end{figure}

The DCDD model clearly produces a resonance when the effects of the
Coulomb barrier are sufficiently strong to confine the wavefunctions.
The asymmetry in the line profiles in figure \ref{fig:over_V} is
caused by the decreasing effect of the Coulomb barrier as the energy
is raised. An interesting feature can be seen in figure
\ref{fig:over_E} that compares overlaps for three different initial
neutron separation energies, 5 keV, 500 keV and 5 MeV. For final state
energies around 200 keV the wavefunction gets above the Coulomb
barrier at about 29 fm and will start oscillating, i.e.\ change sign
periodically. For the case of 5 keV separation energy the initial wave
function will reach out to large distances and the opposite sign
contributions are sufficiently strong to give a clearly visible
interference dip in the upper tail of the resonance.  The finite
energy resolution inherent in the direct decay model gives problems
for resonance structures at low energy since one cannot be sure to
cover the resonance if its width becomes too small and the total
overlap from the model becomes prone to numerical uncertainties and
therefore unreliable.  One can still find the resonance position and
width from the procedure outlined in \cite{Fed09}, but care must be
taken when extracting overlaps.

\begin{figure}[thb]
\centering
     \includegraphics[width=9.cm]{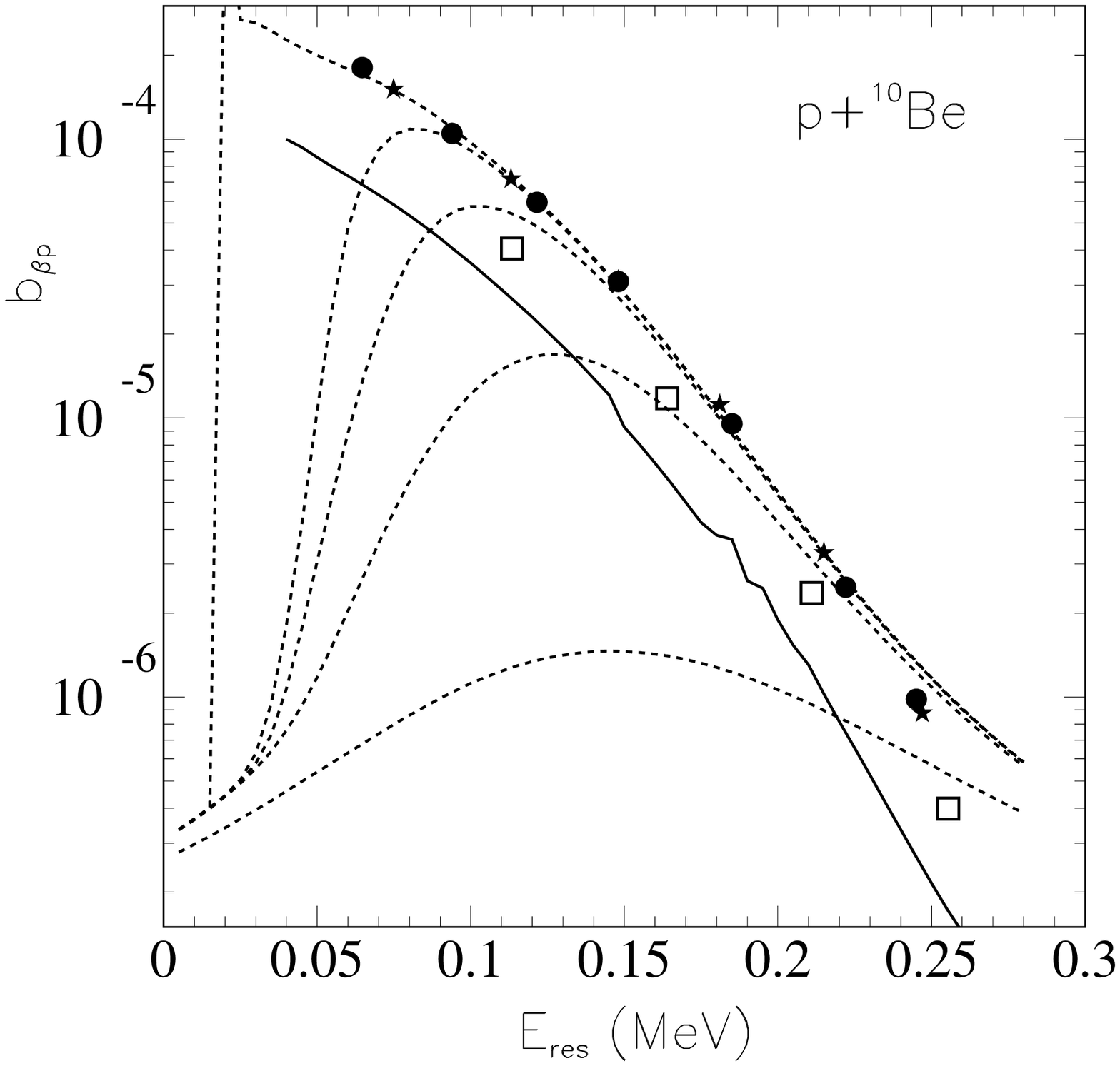} 
     \caption{The calculated branching ratio for decay into
       p+$^{10}$Be as a function of resonance position.
       The stars mark results of DCDD calculations with
       square well potentials. The filled circles are results from
       Woods-Saxon potentials when both inital and final wavefunction
       have one node inside the potential, while the squares
       correspond to no final state nodes. The dashed curves arise
       from integrating the R-matrix expression in eq.\
       (\protect\ref{eq:Rmatrix}) for assumed alpha-decay widths (in
       keV) of 0, 0.1, 1, 10 and 100. The
       full curve is for the expression in eq.\ (\ref{eq:Rshift}).}
\label{fig:branch} 
\end{figure}

Since all involved states have low energy one would expect the results
to be insensitive to the details of the potential shape. This has been
tested by performing calculations also with a Woods-Saxon potential
with parameters taken from the potential used in \cite{Bay11}. Very
similar results were obtained as shown in figure \ref{fig:branch} that
displays the calculated total branching ratio for beta-delayed proton
emission in the two models as a function of the position of the
resonance. The $B_{GT}$ was taken as 3, the sum-rule value for a
single neutron. To obtain the branching ratio the calculated total
decay rate is divided by $w_{tot} = \ln 2/t_{1/2}$. If one has a
mismatch between the nodal structure, e.g.\ one node in the neutron
wave function within the potential and no nodes in the final state
wave functions, the decay rate decreases as also shown in the
figure. (One expects the neutron and proton in the $^{11}$Be decay to
have the same nodal structure.)

The DCDD model calculations point to a resonance dominated decay, so
it is natural to employ also R-matrix calculations of the decay. 
At first I assume
that the level at energy $E_0$ fed in beta-decay only decays via
proton emission. The expression for the decay rate is then (converted
into a differential branching ratio):
\begin{equation}  \label{eq:Rshift}
     \frac{\mathrm{d}b}{\mathrm{d}E} = t_{1/2} \frac{g_A^2}{K} B_{GT}
         \frac{P\gamma^2/\pi}{(E_0+\Delta-E)^2+(P\gamma^2)^2} f_{\beta}(Q-E)\,,
\end{equation}
where $\Delta = -(S(E)-S(E_0))\gamma^2$ and $P$ and $S$ are the
penetrability and shift factors. The value of $\gamma^2$ is taken as
$\hbar^2/(ma^2)$ which is the maximum possible, the
Wigner limit, and where the channel radius used is $a = 1.4
(1+10^{1/3})$ fm. The $B_{GT}$ is again taken as 3.
A simpler approximation sometimes used \cite{Nym90}
is to neglect the energy dependence of the shift factor. Doing this
and allowing also for $\alpha$-decay from the level
gives the following expression:
\begin{equation}  \label{eq:Rmatrix}
     \frac{\mathrm{d}b}{\mathrm{d}E} = t_{1/2} \frac{g_A^2}{K} B_{GT}
         \frac{\Gamma_p/2\pi}{(E_0-E)^2+\Gamma_{tot}^2/4} f_{\beta}(Q-E)\,,
\end{equation}
where $\Gamma_{tot} = \Gamma_{\alpha}+\Gamma_p$, the $\alpha$ decay
width $\Gamma_{\alpha}$ is assumed to be constant over the Q-window
(this should be a good approximation as the $\alpha$ threshold is more
than 2.5 MeV lower), $\Gamma_p = 2P\gamma^2$ and $P$ is the standard
(energy-dependent) penetrability factor. Integration over the Q-window
gives the total branching ratios shown in figure \ref{fig:branch} as a
function of $E_0$ for different values of $\Gamma_{\alpha}$. The
branching ratios agree well with the ones from the DCDD model in the
region where proton decay dominates, note that the calculation with
zero $\Gamma_{\alpha}$ becomes unreliable at very low resonance
energies.

\begin{figure}[thb]
\centering
     \includegraphics[width=9.cm]{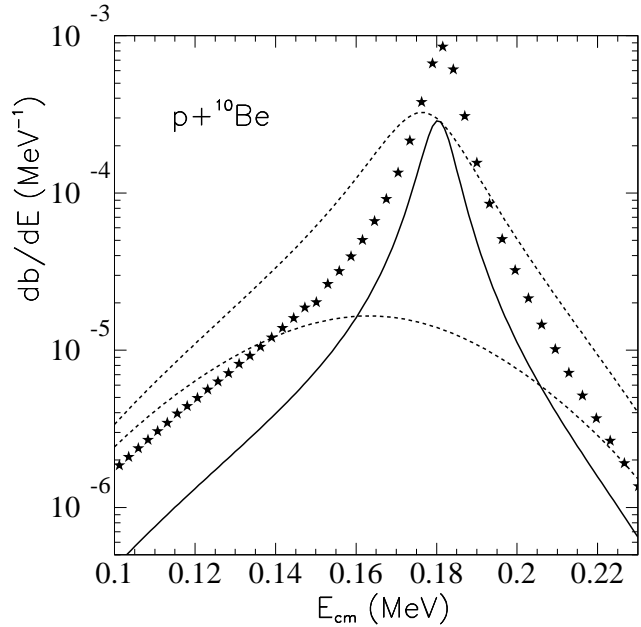} 
     \caption{The energy spectrum for decays into p+$^{10}$Be as a
       function of centre-of-mass energy.
       The stars mark results of DCDD calculations with
       square well potentials. The solid curve are results from
       an R-matrix calculation employing eq.\
       (\protect\ref{eq:Rshift}), the dashed curves arise from eq.\ 
      (\protect\ref{eq:Rmatrix}) for assumed alpha-decay widths
       of 0 keV and 100 keV. See the text for details.}
\label{fig:spec} 
\end{figure}

As discussed in detail in Appendix A R-matrix parameters should not be
identified immediately with experimentally observed quantities;  a
correction factor $(1+\gamma^2 \mathrm{d}S/\mathrm{d}E)^{-1}$ that
for our case decreases slightly from 1/2.5 at 50 keV to 1/3 at 200 keV
enters often. To illustrate this explicitly figure \ref{fig:spec} displays
the differential spectra for the DCDD model with potential depth 33.5
MeV and R-matrix calculations from eqs (\ref{eq:Rshift}) and
(\ref{eq:Rmatrix}) where in both cases $\gamma^2$ is taken as the
Wigner limit, $B_{GT}$ is 3 and $E_0$ is 181 keV (the approximate
resonance position for the DCDD calculation). Comparing first the full
R-matrix calculation, eq.\ (\ref{eq:Rshift}), with the DCDD results
the resonance width and overalll shape are very similar but the
overall strength is reduced by the above factor. This explains
immediately the corresponding reduction in intensity in figure
\ref{fig:branch}. For the simpler expression from eq.\
(\ref{eq:Rmatrix}) the width is clearly too large, again due to the
same correction factor; this underlines that one must
insert the observed width (and not the R-matrix width) when
using eq.\ (\ref{eq:Rmatrix}). A further effect then enters as the energy
dependence of the beta-decay $f$-factor distorts the spectral shape
and moves the peak position several keV down. This effect increases
with $\Gamma_{tot}$ as also seen in figure \ref{fig:spec}, a simple
evaluation where the energy dependence of the width is neglected and
the $f$-factor is approximated as $(Q-E_0)^5$ gives a shift of 
$-(5/8)\Gamma_{tot}^2/(Q-E_0)$.
For the calculation of the total branching ratio a wrong value of the
width does not matter as long as the resonance is narrow since the
integration over the Breit-Wigner then gives a constant, but it could
explain that the results from eq.\ (\ref{eq:Rmatrix}) lie above the
other results in figure \ref{fig:branch} at energies above 250 keV. 
When the correct parameter
values are inserted in the R-matrix calculations the spectra
corresponding to figure \ref{fig:spec} agree very well, as do the
integrated intensities.

\subsection{Implications for R-matrix fits}
If fits are made only in a small energy range it does not matter which
approximation of R-matrix is used. The larger the energy range, and
the larger the effect of having several levels and/or several decay
channels, the more obvious is the need to employ the full theory.
However, whatever method is used, it is essential to distinguish
clearly observed parameters from R-matrix parameters. The main
difference in fitting comes from including or neglecting the shift
factor (compare eqs (\ref{eq:Rshift}) and (\ref{eq:Rmatrix})), whereas
the energy dependence of the level width $\Gamma = 2 P(E) \gamma^2$
may be inserted or not according to whether its variation is
significant. In eq.\ (\ref{eq:Rmatrix}) observed parameters must be
inserted (except for the explicit energy variation of the level
width), in eq. (\ref{eq:Rshift}) the R-matrix parameters. The
conversion between the two parameter sets is, for narrow levels, via
the correction factor $(1+\gamma^2 \mathrm{d}S/\mathrm{d}E)^{-1}$. For
wide levels it eventually becomes meaningless to attempt a conversion.
The important fact to note is now that the Wigner limit applies to the
R-matrix parameter value, whereas the Gamow-Teller sum rule applies to
the observed value.  Note further that the ``observed $B_{GT}$'' only
represents the strength present close to the peak. It may be an
acceptable value for narrow peaks, but for broader peaks one should
apply eq. (\ref{eq:condef}). This holds in particular when
interference occurs.

In some cases the observed $B_{GT}$ gives a misleading impression even
for narrow levels. This is when the small width $\Gamma = 2 P \gamma^2$
is due to a small penetrability rather than a small value for
$\gamma^2$ that measures the strength of the coupling to the outgoing
channel. In this case one may get a sizable contribution also at
higher energies where the penetrability has increased, the ``ghost
peak'' of Barker and Treacy \cite{Bar62}. This effect is also seen in the
$^{11}$Be($\beta$p) case in the DCDD model for resonance energies
below about 65 keV. The effect is mainly due to the change in
penetrability, the change in the shape of the final state wave
function inside the potential that will be present in the DCDD model
is small.

Barker preferred initially \cite{Bar69} to work with ft-values rather
than $B_{GT}$ values. There is no conceptual advantage in doing so for
broad levels, but it may be slightly simpler experimentally and one
does not have to worry about the unfortunate ambiguity in the
literature on whether the ratio of coupling constants $(g_A/g_V)^2$ is
included in $B_{GT}$ or not. For the cases where it is imperative to
use R-matrix fits rather than simply treating a resonance as a bound level
one has no advantage in using ft-values and the comparison to theory
anyway becomes less direct.

Several values of $B_{GT}$ given in the literature will be affected if
the present recommendation is followed. The intriguing case of the
large asymmetry in the decays of the mirror nuclei $^9$Li and $^9$C
will certainly be affected, but the correction factors quoted in
\cite{Pre03} are too small to solve the problem.  An extreme case with
a clear effect is the beta-decay into states in $^{12}$C above the
three-alpha particle threshold. Recent experiments \cite{Hyl09,Hyl10}
gave a summed strength to identified states up to and including the
12.71 MeV state of 1.0--1.1 for $^{12}$B and $^{12}$N.  However,
further strength is clearly seen at 15--16 MeV excitation energy and
if this is interpreted as due to a tail from higher-lying states, as
would be naturally assumed from the R-matrix fitting in \cite{Hyl10},
one would violate the sum rule drastically.  If on the other hand the
observed strength is summed bin by bin, as done in \cite{Hyl09}, one
obtains a strength of about 0.6 for $^{12}$N, which is perfectly
allowed.

\section{Summary}
Using the Gamow-Teller sum rule as a guideline this paper puts forward
a simple general definition of the beta strength in beta-delayed
particle decays. The strength defined in this way differs from the
strength entering in the R-matrix formalism, for extreme cases such as
the $^{12}$N decay the difference is large. When beta decays proceed
through narrow resonances the expression $B^R_{GT}/(1+\gamma^2
\mathrm{d}S/\mathrm{d}E)$ is a good approximation for the summed
strength close to the resonance, but if broad levels are involved or
the $Q_{\beta}$-window is large one should use the definition in eq.\
(\ref{eq:def}) directly.

The suggested definition also differs quantitatively from the ones
employed so far in calculations of decays directly to the continuum,
but has the following advantages: \emph{(i)} the expression for the
decay rate, equation (\ref{eq:def}), is independent of the
normalization of the wavefunctions, \emph{(ii)} the sum over bound
states in the sum-rule is extended as
\begin{equation}
   \sum B_{GT} + \int B_{GT}(E) \mathrm{d}E
\end{equation}
(with the definitions used so far one had to include in  the integral
an explicit phase space factor) and \emph{(iii)} the experimental
determination of beta strength becomes the same independently of
whether the beta decay feeds bound states or the continuum: one simply
uses eq.\ (\ref{eq:def}) for each bin.

Finally, since most theoretical calculations so far have been carried
out in a bound state approximation, any comparison between experiment
and theory must be done with great care once the effects treated in
this paper are significant.

\medskip
\textbf{Acknowledgements}
I would like to thank D.V.\ Fedorov, H.O.U.\ Fynbo, A.S.\ Jensen,
C.A.\ Diget and S.\ Hyldegaard for helpful discussions.

\appendix
\section{R-matrix formalism for beta-delayed decays}

The basic equations for the case of beta-delayed decays are given by
Barker and Warburton \cite{Bar88} and are applicable for several decay
channels and several intermediate levels. In their notation the decay
rate for channel $c$ is
\begin{equation}
  w_c(E) = C^2 f(Q-E) P_c(E) |\sum_{\lambda \mu} g_{\lambda,GT}
  \gamma_{\mu c} A_{\lambda \mu}(E) |^2  \;,
\end{equation}
where the normalization constant $C$ is determined later in their paper,
$f$ is the usual beta decay phase space factor, the penetrability
(shift factor and boundary parameter) for the channel is $P_c$ ($S_c$ and
$B_c$), the levels are indexed by $\lambda$ and $\mu$ with level
energy parameter $E_{\lambda}$, coupling parameter 
$\gamma_{\lambda c}$ to the decay channel $c$ and beta decay parameter
to the level $g_{\lambda,GT}$ corresponding to the parameter in the
main text $B_{GT}^R  = g_{\lambda,GT}^2$. Finally, the elements of the
level matrix are defined through
\begin{equation}
  (A^{-1})_{\lambda \mu} (E) = (E_{\lambda}-E)\delta _{\lambda \mu} -
   \sum_c (S_c-B_c+iP_c)\gamma _{\lambda c} \gamma _{\mu c} \;.
\end{equation}
For a single isolated level and a single decay channel for the levels
this expression reduces to the one given just before eq.\ (\ref{eq:rho}) above.
The standard Breit-Wigner expression is obtained by approximating
$2P(E_{\lambda}) \gamma^2_{\lambda}$ by $\Gamma$ and
$E_{\lambda}+(S(E_{\lambda})-B) \gamma^2_{\lambda}$ by $E_0$. Brune
\cite{Bru02} has provided an altervative parametrization that avoids
boundary parameters and may be more useful in applications.

Barker has argued \cite{Bar69,Bar88,Bar96} that the best approximation
to the integral of $\rho$ (given in eq. (\ref{eq:rho})) is
\begin{equation}  \label{eq:Barker}
  \int_0^{\infty} \rho_{\lambda}(E) \mathrm{d}E = \frac{\pi}{
       1+\gamma^2_{\lambda} \frac{\mathrm{d}S}{\mathrm{d}E}|_{E_{\lambda}}}
\end{equation}
and that the deduced $B_{GT}$ (similar to extracted $\Gamma$-values)
must be reduced by dividing with the same denominator. The argument
comes from Lane and Thomas \cite{Lan58} and is based on making a first
order Taylor expansion of $S(E) = S(E_{\lambda}) +
(E-E_{\lambda})\mathrm{d}S/\mathrm{d}E$. A standard Breit-Wigner form
can then be obtained by pulling out the square of the factor 
$F = 1+\gamma^2_{\lambda}\frac{\mathrm{d}S}{\mathrm{d}E}|_{E_{\lambda}}$
from the energy difference in the denominator in eq.\
(\ref{eq:Rshift}) so that the effective width,
often called observed width, becomes $\Gamma/F$ and similar for the
$B_{GT}$ value. 
Essentially, in this approximation the integral is performed as if the
shape in the peak region can be extended trivially to all energies.

A more accurate way of evaluating the total integral is as follows.
The expression for $\rho(E)$ can be analytically continued to complex
values of the energy $E$ if one places a cut just above the negative
real axis, this is needed since the penetrability contains a factor
$kR =\sqrt{c'E}$.  The poles of $\rho_{\lambda}(E)$ appear for
\[  E = E_{\lambda}+(S(E)-B) \gamma^2_{\lambda} \pm i P(E)\gamma^2_{\lambda}
\]
and apart from the ``classical'' pole at $E_-=E_0-i\Gamma/2$ (and its
mirror $E_+=E_0+i\Gamma/2$) one can have poles from the terms 
$P(E)$ and $S(E)$. One now extends
the integral along the negative real axis and closes it in the lower
half so that the residue theorem can be used. The added part in the
lower complex plane gives a vanishing contribution and the integral
along the negative real axis gives a purely imaginary contribution
(except for possible poles on the axis which then have to be evaluated
explicitly) due to the factor $\sqrt{E}$, this can be shown explicitly
for neutral particles but probably holds in general.  The integral can
therefore be evaluted simply from the residues at the poles of the
function in the lower half where the main contribution (at least for
small $\gamma^2$) will be from the pole at $E_-$. 
Close to this pole the behaviour is
similar to that of the standard Breit-Wigner function but the residue,
which for the Breit-Wigner is $2\pi i \Gamma/2 / (E_+-E_-) = 2\pi i
\Gamma/2 / (2i\Gamma/2) = \pi$,
becomes more complex due to the analytic continuation of
$P(E)$ into the complex plane. The correct result for the integral
turns out to be $\pi$ also for s-wave neutrons (where $P(E)=kR$ and
$S(E) = 0$) but for higher angular momenta deviations will
occur. One can show in general that each pole $E_p$ will give a
contribution to the integral of
\[  \pi \left( \pm \left[1 - \gamma^2_{\lambda} \left.
  \frac{\mathrm{d}S}{\mathrm{d}E}\right|_{E_p} \right]  \left.
     - i \gamma^2_{\lambda} \frac{\mathrm{d}P}{\mathrm{d}E}\right|_{E_p}
   \right)^{-1} \;,
\]
i.e., if the imaginary term can be neglected to lowest order, at
first glance a similar result to eq.\ (\ref{eq:Barker}) but with opposite sign!
Furthermore, the contributions from the poles that arises from $P(E)$ give
a contribution that is of the same order as the correction terms from the
classical pole. In fact, an explicit calculation for p-wave neutrons
(where $P(E) = (c'E)^{3/2}/(1+c'E)$ and $S(E) = -1/(1+c'E)$)
gives to lowest order that the integral is
\[  \pi \left( 1 + 2\frac{c' \gamma^2_{\lambda}}{(1+c'
    E_{\lambda})^2} \right)
\]
%\[  \pi \left( 1 + \frac{5}{2}\frac{c' \gamma^2_{\lambda}}{(1+c'
%    E_{\lambda})^2} + \frac{1}{2}\frac{c' \gamma^2_{\lambda}}{(1+c'
%    E_{\lambda})^3}  \right)
%\]
where the classical pole and a pole close to $-1$ contributes equally
to the last term.  The reason that the Lane and Thomas expansion gives
the wrong result is, as pointed out explicitly in \cite{Hyl10} for the
case of the Hoyle state, that part of the strength appears away from
the pole position as so-called ghosts \cite{Bar62}.  Whether this
contribuiton is physically important or not depends on the specific
circumstances of a decay. The summed strength close to the resonance
energy will for narrow levels be given to a good approximation by eq.\
(\ref{eq:Barker}).

For the one-level multi-channel case one can see easily from the
general formulae in \cite{Bar88} that the only change is to insert a
sum over all channels in the terms $(S(E)-B)\gamma^2$ and
$P(E)\gamma^2$. Summing the contribution in all outgoing channels one
therefore obtains a result similar to the one-level one-channel
case. However, in the multi-level one-channel case interference
between the levels will appear. Numerically this does not seem to have
a major effect; in fact, for levels that interact constructively
between the two peak positions it seems that the total areas are less
perturbed than for individual single levels even for cases with rather
strong interference (the desctructive interference away from the
levels seems to suppress the contributions that makes integrals deviate
from $\pi$).


\begin{thebibliography}{99}
\bibitem{Bla08}B. Blank and M.J.G. Borge, Prog. Part. Nucl. Phys.
  {\bf 60} (2008) 403
\bibitem{Pfu12}M. Pf\"{u}tzner, M. Karny, L.V. Grigorenko and
  K. Riisager, Rev. Mod. Phys. {\bf 84} (2012) 567
\bibitem{Bar69}F.C. Barker, Aust. J. Phys. {\bf 22} (1969) 293
\bibitem{Bar88}F.C. Barker and E.K. Warburton, Nucl. Phys. {\bf A487}
  (1988) 269
\bibitem{Nym90}G. Nyman et al., Nucl. Phys. {\bf A510} (1990) 189
\bibitem{Bar96}F.C. Barker, Nucl. Phys. {\bf A609} (1996) 38
\bibitem{Ber68}T. Berggren, Nucl. Phys. {\bf A109} (1968) 265
\bibitem{Ber93}T. Berggren and P. Lind, Phys. Rev. {\bf C47} (1993) 768
\bibitem{Ber73}T. Berggren, Phys. Lett. {\bf B44} (1973) 23
\bibitem{Rom75}W.J. Romo, Nucl. Phys. {\bf A237} (1975) 275
\bibitem{Myo98}T. Myo, A. Ohnishi and K. Kato, Prog. Theor. Phys. {\bf 99} (1998) 801
%\bibitem{Myo97}T. Myo and K. Kato, Prog. Theor. Phys. {\bf 98} (1997) 1275
%\bibitem{Civ08}O. Civitarese, R.J. Liotta and M.E. Mosquera,
%  Phys. Rev. {\bf C78} (2008) 064308
\bibitem{Mic09} N. Michel, W. Nazarewicz, M. P{\l}oszajczak and
  T. Vertse, J. Phys. {\bf G36} (2009) 013101
\bibitem{Rii90}K. Riisager et al., Phys. Lett. {\bf 235} (1990) 30
\bibitem{Zhu93} M.V. Zhukov, B.V. Danilin, L.V. Grigorenko and
  N.B. Shul'gina, Phys. Rev. {\bf C47} (1993) 2937
\bibitem{Bay94}D. Baye, Y. Suzuki and P. Descouvemont,
  Prog. Theor. Phys. {\bf 90} (1994) 271
\bibitem{Tur06}E.M. Tursunov, D. Baye and P. Descouvemont, Phys. Rev. {\bf C73}
 (2006) 014303
\bibitem{Bar94}F.C. Barker, Phys. Lett. {\bf B322} (1994) 17
\bibitem{Lan58}A.M. Lane and R.G. Thomas, Rev. Mod. Phys. {\bf 30}  (1958) 257
\bibitem{Des10} P. Descouvemont and D. Baye, Rep. Prog. Phys. {\bf 73}
  (2010) 036301
\bibitem{Rob75}D. Robson, in Nuclear Spectroscopy and Reactions,
  ed. J. Cerny (Academic, New York, 1975), vol. D, p. 179
\bibitem{Bor13} M.J.G. Borge et al., J. Phys. {\bf G40} (2013) 035109
\bibitem{mas12} M. Wang et al., Chinese Phys. C \textbf{36} (2012) 1603
\bibitem{Bay11} D. Baye and E.M. Tursonov, Phys. Lett. {\bf B696}
  (2011) 464
\bibitem{Zhu95} M.V. Zhukov, B.V. Danilin, L.V. Grigorenko and
  J.S. Vaagen, Phys. Rev. C \textbf{52} (1995) 2461
\bibitem{Fed09} D.V. Fedorov, A.S. Jensen and M. Th{\o}gersen,
  Few-Body Syst. \textbf{45} (2009) 191
\bibitem{Bar62} F.C. Barker and P.B. Treacy, Nucl. Phys. {\bf 38}
  (1962) 33
\bibitem{Pre03} Y. Prezado et al., Phys. Lett. {\bf B576} 55
\bibitem{Hyl09}S. Hyldegaard et al., Phys. Lett. {\bf B678} (2009) 459
\bibitem{Hyl10}S. Hyldegaard et al., Phys. Rev. {\bf C81} (2010) 024303
\bibitem{Bru02} C.R. Brune, Phys. Rev. {\bf C66} (2002) 044611
\end{thebibliography}
\end{document}